\documentclass[twocolumn,prx,floatfix,superscriptaddress]{revtex4-1}
\usepackage{amsfonts,amsmath,amssymb,color,times,graphicx}
\definecolor{darkblue}{rgb}{0, 0, 0.8}
\usepackage[colorlinks=true, breaklinks=true, linkcolor=blue, citecolor=darkblue, urlcolor=darkblue]{hyperref}
\usepackage{graphicx}
\usepackage{bmpsize}
\usepackage{amsfonts}
\usepackage{gensymb}
\usepackage{bm}
\usepackage{braket}
\usepackage{mathtools}

\begin{document}

\title{Doping dependence of the magnetic ground state in the frustrated magnets Ba$_2$$\textit{\textbf{M}}$Te$_{1-x}$W${_x}$O$_6$ ($\textit{\textbf{M}}$ = Mn, Co)}
\author{Chaoxin Huang}
\affiliation{Center for Neutron Science and Technology, Guangdong Provincial Key Laboratory of Magnetoelectric Physics and Devices, School of Physics, Sun Yat-sen University, Guangzhou, 510275, China}
\author{Lisi Li}
\affiliation{Center for Neutron Science and Technology, Guangdong Provincial Key Laboratory of Magnetoelectric Physics and Devices, School of Physics, Sun Yat-sen University, Guangzhou, 510275, China}
\author{Peiyue Ma}
\affiliation{Center for Neutron Science and Technology, Guangdong Provincial Key Laboratory of Magnetoelectric Physics and Devices, School of Physics, Sun Yat-sen University, Guangzhou, 510275, China}
\author{Xing Huang}
\affiliation{Center for Neutron Science and Technology, Guangdong Provincial Key Laboratory of Magnetoelectric Physics and Devices, School of Physics, Sun Yat-sen University, Guangzhou, 510275, China}
\author{Tao Xie}
\thanks{Corresponding author: xiet69@mail.sysu.edu.cn}
\affiliation{Center for Neutron Science and Technology, Guangdong Provincial Key Laboratory of Magnetoelectric Physics and Devices, School of Physics, Sun Yat-sen University, Guangzhou, 510275, China}
\author{Meng Wang}
\thanks{Corresponding author: wangmeng5@mail.sysu.edu.cn}
\affiliation{Center for Neutron Science and Technology, Guangdong Provincial Key Laboratory of Magnetoelectric Physics and Devices, School of Physics, Sun Yat-sen University, Guangzhou, 510275, China}

\begin{abstract}
Theoretically, the relative change of the Heisenberg-type nearest-neighbor coupling $J_1$ and next-nearest-neighbor coupling $J_2$ in the face-centered-cubic lattice can give rise to three main antiferromagnetic orderings of type-I, type-II, and type-III. However, it is difficult to tune the $J_2/J_1$ ratio in real materials. Here, we report studies on the influence of Te$^{6+}$ and W$^{6+}$ ions replacement to the magnetic interactions and the magnetic ground states in the double-perovskite compounds Ba$_2$$M$Te$_{1-x}$W$_{x}$O$_6$ ($M$ = Mn, Co). For Ba$_2$MnTe$_{1-x}$W$_{x}$O$_6$, the W$^{6+}$ doping on Te$^{6+}$ site is successful in $0.02 \leq x \leq 0.9$ with short-range orders of the type-I ($0.02 \leq x \leq 0.08$) and type-II ($0.1 \leq x \leq 0.9$). In Ba$_2$CoTe$_{1-x}$W${_x}$O$_6$, x-ray diffraction measurements reveal two crystal structures, including the trigonal phase ($0 \leq x \leq 0.1$) and the cubic phase ($0.5 \leq x \leq 1$), between which is a miscibility gap. Two magnetic transitions are identified in the trigonal phase due to two magnetic subsystems, and the type-II magnetic order is observed in the cubic phase. Magnetic phase diagrams of Ba$_2M$Te$_{1-x}$W$_{x}$O$_6$ ($M$ = Mn, Co) are established. Our work shows that the magnetic interactions and ground states of Ba$_2$$M$Te$_{1-x}$W$_x$O$_6$ can be tuned effectively by the replacement of Te$^{6+}$ by W$^{6+}$ ions.
\end{abstract}
\maketitle

\section{Introduction}

\begin{figure*} [ht]
\centering
\includegraphics[width=1\linewidth]{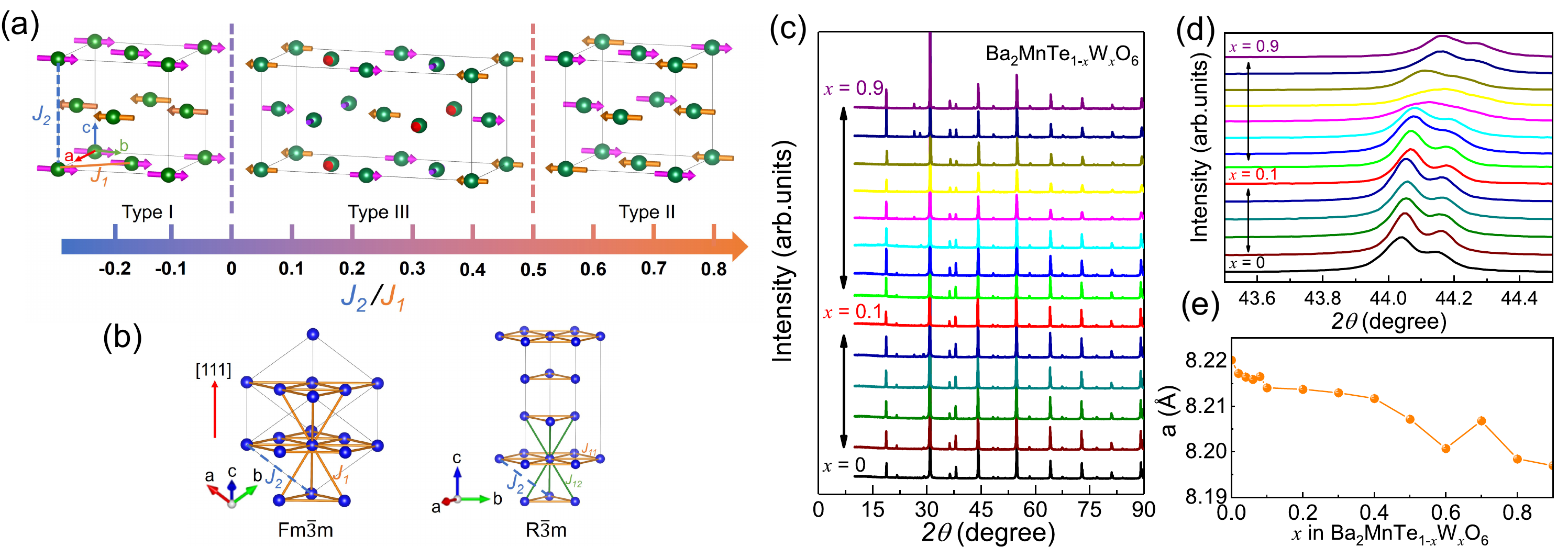}
\caption{(a) The magnetic structures of type-I, type-II, and type-III with different ratios of $J_2/J_1$ based on the $J_1$-$J_2$ model in the FCC lattices\cite{Sun2018}. (b) Crystal structure of Ba$_{2}$MnTeO$_{6}$ only showing Mn atoms with cubic (left)\cite{Mustonen2020} and trigonal (right)\cite{Li2020} descriptions. The (111) plane, as shown on the left of the cubic lattice, is equivalent to the $ab$ plane of the trigonal lattice, where the Mn atoms form triangular layers. (c) The XRD patterns and (d) the local magnification of Ba$_2$MnTe$_{1-x}$W$_x$O$_6$. (e) Doping dependence of the lattice constant $a$ of Ba$_2$MnTe$_{1-x}$W$_x$O$_6$ refined from the XRD results.}
\label{Figure1}
\end{figure*}

Geometrically frustrated materials have been widely studied to explore novel magnetic ground states such as non-collinear magnetic order, quantum spin ice, and quantum spin liquids\cite{Balents2010,Zhou2017,Savary2017,Chamorro2021}.
In two-dimensional systems, geometrically frustrated magnets always hold triangular and kagome lattices constructed by the edge- and corner-sharing triangles, respectively. While the counterparts in three-dimensional cases are face-centered-cubic (FCC) and pyrochlore lattices with edge- and corner-sharing tetrahedra, respectively\cite{Ramirez1994,Stockert2020}. The FCC lattice with nearest-neighbor (NN) interaction $J_1$ and next-nearest-neighbor (NNN) interaction $J_2$ has been studied extensively from the theoretical side\cite{Tahir-Kheli1966,Sun2018}. When $J_1$ is antiferromagnetic (AFM), an FCC system can have three different magnetic ground states depending on the ratio of $J_2/J_1$, including type-I ($J_2/J_1$$<$0) with alternating ferromagnetic layers along [001], type-II ($J_2/J_1$$>$0.5) with alternating ferromagnetic layers along [111], and type-III (0$<$$J_2/J_1$$<$0.5) with alternating ferromagnetic layers along [210], as shown in Fig. \ref{Figure1}(a)\cite{Rausch2011,Sun2018}. However, realizing a tunable magnetic ground state via changing the ratio of $J_2/J_1$ is still rare.

Chemical doping is an effective method to manipulate the superexchange magnetic interactions and the magnetic ground states of magnetic systems, and it can induce novel properties like unconventional superconductivity\cite{Stewart2017}. Te$^{6+}$ and W$^{6+}$ ions replacement is a good example of chemical dopant in quantum frustrated magnets\cite{Zhu2014,Mustonen2018,Mustonen2018a}. For example, the magnetic ground state of Cr$_{2}$(Te$_{1-x}$W$_{x}$)O$_{6}$ evolves from a single-layer AFM order to a bilayer AFM order with the increasing doping $x$\cite{Zhu2014}. In Sr$_2$CuTeO$_6$, an $S$ = 1/2 square-lattice antiferromagnet, the N\'{e}el AFM structure can be quickly tuned to a short-range order by W$^{6+}$ doping\cite{Babkevich2016,Mustonen2018,Mustonen2018a,Watanabe2018,Yoon2021}, and a novel random singlet phase emerges with $0.2\leq x \leq 0.6$~\cite{Hong2021,Hu2021,Fogh2022}. At the end point of the W$^{6+}$ doping, the magnetic ground state of Sr$_2$CuWO$_6$ is a columnar AFM order~\cite{Walker2016}. However, such doping effects are rarely studied in the FCC magnetic frustrated system.

Double perovskite compounds Ba$_2$$M$TeO$_6$ ($M$ = Mn, Co) with FCC lattice provide a platform to study the doping effects and novel physics in the FCC system.
The lattice structure of Ba$_2$MnTeO$_6$ is still not well determined, and two structural models, including trigonal (space group: $R\bar{3}m$) and cubic (space group: $Fm\bar{3}m$) lattices are proposed \cite{Wulff1998,Li2020,Khatua2021,Mustonen2020}, as shown in Fig. \ref{Figure1}(b). These two cases have a similar energy scale of magnetic interactions, and both support the type-I magnetic structure\cite{Mustonen2020,Li2020}. Ba$_2$CoTeO$_6$ holds a trigonal structure with the space group $P\bar{3}m$ and Co$^{2+}$ cations form two magnetic subsystems including a Heisenberg-like antiferromagnet with triangular lattice and an Ising-like antiferromagnet with buckled honeycomb lattice~\cite{Ivanov2010,Chanlert2016,Chanlert2017,Kojima2022}. However, Ba$_2$MnWO$_6$ and Ba$_2$CoWO$_6$ both crystallize into a cubic structure ($Fm\bar{3}m$) with the type-II magnetic order~\cite{Mutch2020,Cox1967,Hanna2024}. Therefore, it is very interesting to perform studies in Ba$_2$$M$Te$_{1-x}$W$_x$O$_6$ ($M$ = Mn, Co) to figure out how the magnetic ground states evolve from Ba$_2$$M$TeO$_6$ to Ba$_2$$M$WO$_6$.

In this paper, we present studies on the doping dependence of crystal structure and magnetic ground state in Ba$_2$$M$Te$_{1-x}$W$_x$O$_6$ ($M$ = Mn, Co) by employing x-ray diffraction (XRD), magnetic susceptibility and heat capacity measurements. For Ba$_2$MnTe$_{1-x}$W$_x$O$_6$, the type-I magnetic order is gradually suppressed from $x=0$ to $x=0.08$. With $0.1 \leq x \leq 0.9$, the system enters a type-II short-range magnetically ordered state and finally becomes a long-range type-II order at $x=1$. Ba$_2$CoTe$_{1-x}$W$_x$O$_6$ crystalizes in a triangular-lattice structure with two magnetic phase transitions in the doping region $0 \leq x \leq 0.1$. It holds a cubic structure with a short-range type-II magnetic order at $0.5 \leq x < 1$ until a long-range type-II order forms at $x=1$. When $0.1 < x < 0.5$, the obtained materials are mixtures of Ba$_2$CoTeO$_6$ and Ba$_2$CoWO$_6$, which are separated by a miscibility gap. Based on these results, we construct the magnetic phase diagrams of Ba$_2$MnTe$_{1-x}$W$_x$O$_6$ and Ba$_2$CoTe$_{1-x}$W$_x$O$_6$. The microscopic origin of the doping influences on the magnetic ground states is also discussed.

\section{Methods}
Polycrystalline samples of Ba$_2$$M$Te$_{1-x}$W$_x$O$_6$ ($M$ = Mn, Co) were synthesized by the solid-state reaction method. For Ba$_2$MnTe$_{1-x}$W$_x$O$_6$, the powdered BaCO$_3$ (99.99$\%$), MnCO$_3$ (99$\%$), TeO$_2$ (99.99$\%$) and WO$_3$ (99.99$\%$) were mixed in a molar ratio of 2:1:1$-$$x$:$x$ according to the doping contents, while the reactant is a mixture of BaCO$_3$ (99.99$\%$), Co$_3$O$_4$ (99.99$\%$), TeO$_2$ (99.99$\%$) and WO$_3$ (99.99$\%$) in a molar ratio of 2:$1/3$:1$-$$x$:$x$ for Ba$_2$CoTe$_{1-x}$W$_x$O$_6$. The initial mixtures were ground and pressed into pellets. The pellets were placed in an alumina crucible and then put into a muffle furnace for high-temperature sintering at $1150-1300$ $^{\circ}$C. The target materials were obtained after grinding, pressing, and sintering several times. To determine the structure and purity of the samples, powder XRD experiments were conducted on a Rigaku diffractometer with Cu K$\alpha$$_1$ ($\lambda$ = 1.5406 {\AA}). Direct current (DC) magnetic susceptibility, alternating current (AC) magnetic susceptibility, and heat capacity were measured using the VSM, ACMS, and HC modules on a physical property measurement system (PPMS, Quantum Design), respectively. The data were collected in the temperature range of $2-300$ K and a magnetic field range of $0-14$ T.

\section{Results}
\subsection{B$\textrm{a}_2$M$\textrm{n}$T$\textrm{e}_{1-x}$W$_x$O$_6$}

Powder XRD results of Ba$_2$MnTe$_{1-x}$W$_x$O$_6$ are shown in Fig. \ref{Figure1}(c). It should be noted that the diffraction patterns of the two previously mentioned structural models, trigonal (space group: $R\bar{3}m$) and cubic (space group: $Fm\bar{3}m$) for Ba$_2$MnTeO$_6$\cite{Li2020,Khatua2021,Mustonen2020} only have tiny differences, and cannot be distinguished by XRD measurements. However, both structural models exhibit the FCC Heisenberg physics. Since we mainly focus on the magnetic properties instead of the structural problem in this work, the cubic (space group: $Fm\bar{3}m$) structure is thus used to refine the XRD data. The refinements of two representative doping contents of Ba$_2$MnTe$_{1-x}$W$_x$O$_6$ are summarized in Appendix [Figs.~\ref{PXRD}(a) and (b)]. From the refinements, a small amount of impurities of Ba$_3$TeO$_6$ and BaWO$_4$ are identified. Since the impurities are tiny and nonmagnetic, we will not discuss them. A magnification of the XRD patterns around the (400) peak and the lattice constant $a$ refined from XRD as a function of $x$ is shown in Figs.~\ref{Figure1}(d) and (e). The peaks smoothly shift to a higher angle with the increasing $x$, consistent with the lattice constant $a$ decreasing in Fig. \ref{Figure1}(e).

\begin{figure}
\centering
\includegraphics[width=1\linewidth]{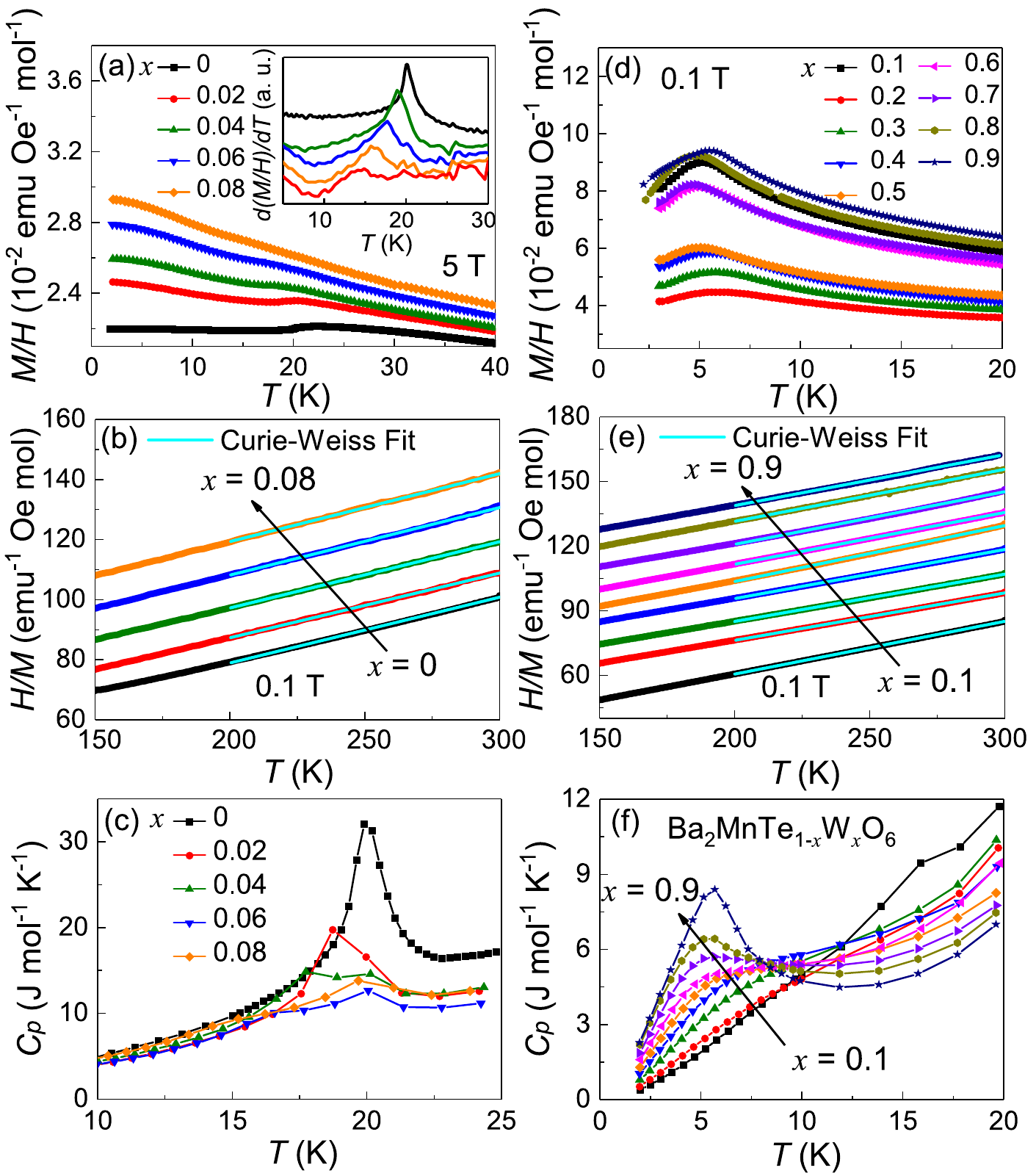}
\caption{(a) Temperature dependence of magnetic susceptibility at 5 T for Ba$_2$MnTe$_{1-x}$W$_x$O$_6$ ($0 \leq x \leq 0.08$). The inset in (a) shows the derivatives of susceptibility curves. (b) CW fittings of the high-temperature susceptibility (above 200 K) of Ba$_2$MnTe$_{1-x}$W$_x$O$_6$ ($0 \leq x \leq 0.08$). These curves have been vertically shifted for clarity. (c) Heat capacity at zero field at doping $0 \leq x \leq 0.08$. (d) Temperature dependence of magnetic susceptibility at 0.1 T for Ba$_2$MnTe$_{1-x}$W$_x$O$_6$ ($0.1 \leq x \leq 0.9$). (e) CW fittings of the high-temperature susceptibility (above 100 K) of Ba$_2$MnTe$_{1-x}$W$_x$O$_6$ ($0.1 \leq x \leq 0.9$). (f) Heat capacity at zero field at doping $0.1 \leq x \leq 0.9$.}
\label{Figure2}
\end{figure}

\begin{figure}
\centering
\includegraphics[width=1\linewidth]{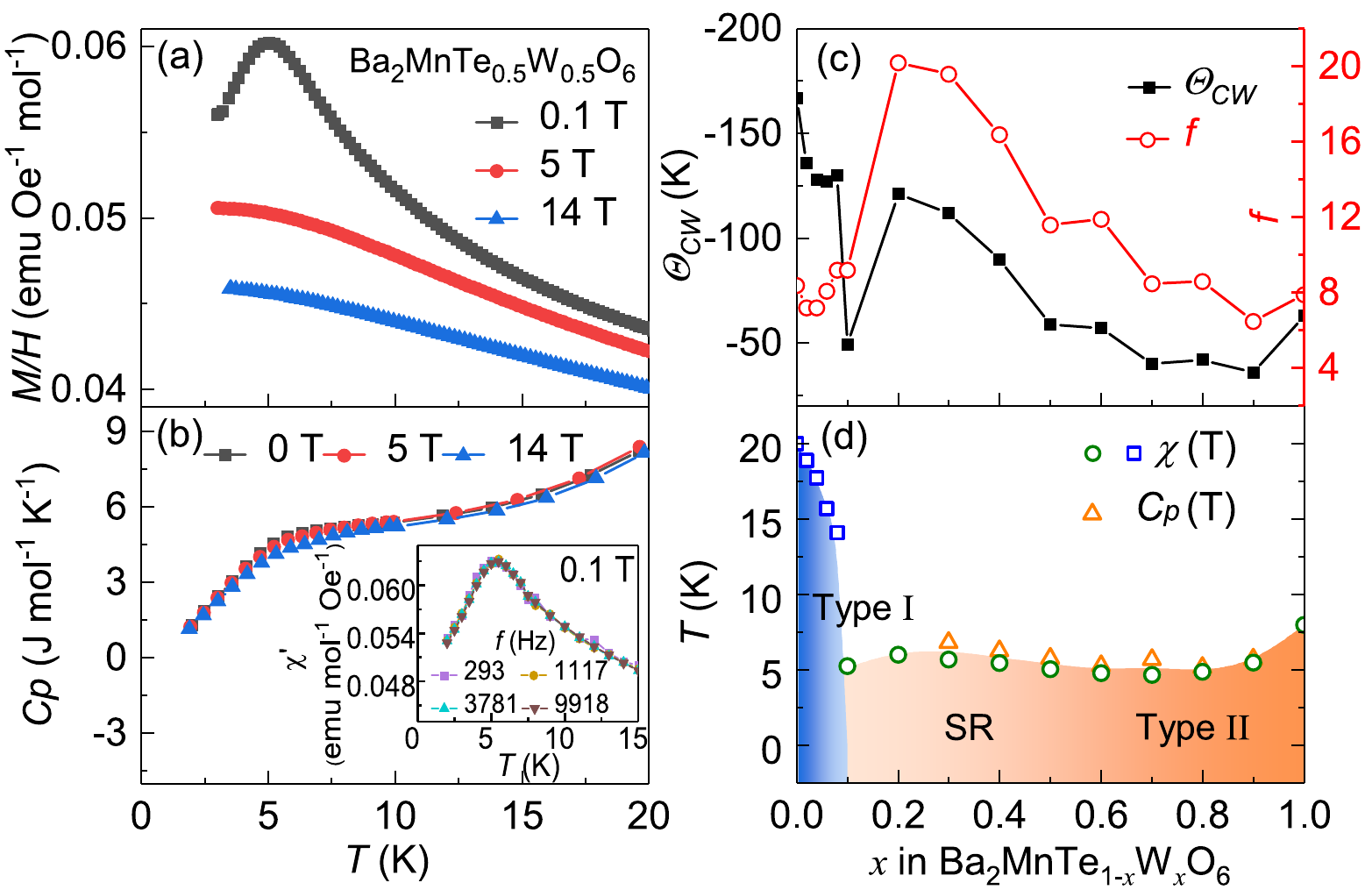}
\caption{(a)Temperature dependence of $M/H$ and (b) heat capacity in different magnetic fields of Ba$_2$MnTe$_{0.5}$W$_{0.5}$O$_6$. The inset in (b) shows the AC susceptibility in different frequencies with a DC magnetic field at 0.1 T. (c) Doping dependence of CW temperature $\mit\Theta_{CW}$ and frustration coefficient $f$. (d) Magnetic phase diagram including type-I (Blue), type-II (Orange), and short-range (SR) AFM order regions. The data at $x = 1.0$ in (c) and (d) is obtained from Ref.~\cite{Mutch2020}.}
\label{Figure3}
\end{figure}

Figure~\ref{Figure2} shows the temperature dependence of DC susceptibility and heat capacity of Ba$_2$MnTe$_{1-x}$W$_x$O$_6$. In the DC susceptibility, the AFM transition temperature ($T_N$) can be obtained by the first derivative of susceptibility, as shown in the insert of Fig.~\ref{Figure2}(a). Together with the heat capacity results in Fig.~\ref{Figure2}(c), the $T_N$ of Ba$_2$MnTeO$_6$ can be determined to be $\approx$ 20 K. The AFM transition is gradually suppressed as the doping level increases from $x = 0$ to 0.08. It should be noted that the weak intensities in heat capacity and the first derivative of susceptibility around 20 K for $x$ = 0.04, 0.06, and 0.08 in Figs.~\ref{Figure2}(a) and (c) are caused by Ba$_2$MnTeO$_6$ impurity, which cannot be observed in XRD refinements. The broad (non-$\lambda$-like sharp) peaks in the heat capacity results for doping $0.02 \leq x \leq 0.08$ compounds indicate the short-range feature of the magnetic order. Similar measurements for the higher doping levels with $0.1 \leq x \leq 0.9$ of Ba$_2$MnTe$_{1-x}$W$_x$O$_6$ are shown in Figs.~\ref{Figure2}(d) to (f). In this doping range, broad peaks around 5 K are observed in DC magnetic susceptibility, which indicates that $T_N$ does not change substantially. In heat capacity, the peaks around 5 K are induced and enhanced with the increasing $x$. There is no sharp peak in heat capacity for $0.1 \leq x \leq 0.8$, suggesting short-range AFM orders below the corresponding $T_N$s. Since the magnetic ground state of Ba$_{2}$MnWO$_{6}$ is type-II magnetic order with $T_N$ = 8 K\cite{Mutch2020}, it is suggested that the magnetic order gradually transforms from a short-range correlation similar to type-II order to a long-range order with the increasing $x$ until $x$ = 0.9 and 1. In Figs.~\ref{Figure2}(b) and (e), the temperature dependence of the inverse magnetic susceptibility and the Curie-Weiss (CW) fittings at high temperatures are shown. The data above 200 K are fitted according to the CW law $\chi= \chi_0 + C/(T-\mit\Theta_{CW})$\cite{StephenBlundell2001}, where $\chi_0$ includes the contribution from diamagnetism and Van Vleck paramagnetism, $C$ is the Curie constant, and $\mit\Theta_{CW}$ is the CW temperature. The $T_{N}$, the fitted $\mit\Theta_{CW}$, the deduced effective magnetic moment $\mu_{eff}$ and the frustrated coefficient $f$ = $-\mit\Theta_{CW}/T_{N}$ are summarized in Table \ref{Table1}. The obtained effective magnetic moments $\mu_{eff}$ are $\sim$5.54-6.3~$\mu_B$, which are close to the theoretical effective magnetic moment $\mu_{eff}$ = $g\sqrt{S(S+1)}$ = 5.91 $\mu_B$ for a spin-only $S$ = $5/2$ system. The negative CW temperatures $\mit\Theta_{CW}$ confirm the dominantly antiferromagnetic interactions in these compounds. As shown in Fig. \ref{Figure3}(c), the $\mit\Theta_{CW}$ decreases with the increasing $x$, indicating that the strength of the magnetic interactions is weakened upon doping. As a result, the frustration coefficient $f$ changes dramatically with $x$ and forms a peak around $x$ = 0.2 [Fig.~\ref{Figure3}(c)]. The large frustration coefficients for the intermediate dopings ($0.2 \leq x \leq 0.6$) suggest strong magnetic frustrations in these doping compounds. Interestingly, a dip of $\mit\Theta_{CW}$ at $x = 0.1$ is deemed the turning point of the two magnetic phases. However, there is no obvious peak in the heat capacity result of $x = 0.1$, which may indicate a magnetically disordered ground state.

\begin{table}
\caption{Effective moments $\mu_{eff}$, CW temperatures $\mit\Theta_{CW}$, $T_{N}$, and frustration coefficient $f$ = $-\mit\Theta_{CW}/T_{N}$ of Ba$_2$MnTe$_{1-x}$W$_{x}$O$_6$ determined by CW fit.}
 \vspace{2mm}
\begin{tabular}{ccccc}
\hline
$x$      & $\mu_{eff}$($\mu_B$) & $\mit\Theta_{CW}$(K) &$T_{N}$(K) &$f$ \\ \hline
0   & 6.05 & -167 &20.0 &8.4 \\
0.02 & 6.06	&-136 &18.9 &7.2 \\
0.04 & 5.99 &-128 &17.7 &7.2 \\
0.06 & 5.94	&-127 &15.7 &8.1 \\
0.08  &5.92	&-130 &14.1 &9.2 \\
0.1  & 5.69 &-49 &5.3 &9.2 \\
0.2  & 6.00	&-121 &6.0 &20.2 \\
0.3  & 6.02	&-112 &5.7 &19.6 \\
0.4  & 5.91 &-90 &5.5 &16.4 \\
0.5   &5.54	&-59 &5.1 &11.6 \\
0.6  & 5.74	&-57 &4.8 &11.9 \\
0.7  & 5.72	&-40 &4.7 &8.5 \\
0.8  & 5.71	&-42 &4.9 &8.6 \\
0.9   & 5.78 &-36 &5.5 &6.5 \\
1.0 \cite{Mutch2020}  & 6.3 &  -63 &8.0 &7.9 \\   \hline \hline
\end{tabular}
\label{Table1}
\end{table}

The magnetic properties of Ba$_2$MnTe$_{0.5}$W$_{0.5}$O$_6$ are further studied by DC magnetic susceptibility and heat capacity at higher fields and AC magnetic susceptibility to investigate its magnetic ground state, as depicted in Figs.~\ref{Figure3}(a) and (b). The broad peak at $\sim$5 K in magnetic susceptibility can be suppressed by the applied magnetic fields, resulting from the destroy of the short-range AFM order and the appearance of a weak ferromagnetic component induced by spin polarization. However, the broad hump around $\sim$5-7 K in heat capacity only changes a little against the magnetic fields, suggesting that the AFM correlations are still robust although the peak corresponding to the short-range order in magnetic susceptibility can be suppressed by magnetic fields. As shown in the inset of Fig. \ref{Figure3}(b), the broad peak around 5 K in the temperature dependence of AC magnetic susceptibility does not shift with frequency, ruling out the spin-freezing state. Therefore, the magnetic ground state of Ba$_2$MnTe$_{0.5}$W$_{0.5}$O$_6$ is considered to be a short-range AFM order.

To summarize, the magnetic ground state for compounds with $0.02 \leq x \leq 0.08$ is a short-range type-I AFM order. In contrast, the magnetic behaviors of samples with doping $0.1 \leq x \leq 0.9$ differ from the compounds at $0 < x < 0.1$ and change systematically with the increasing doping $x$. As $x$ increases further, the magnetic ground state of the system gradually evolves to the long-range type-II order. Based on these results, a magnetic phase diagram of Ba$_2$MnTe$_{1-x}$W$_x$O$_6$ is established, as shown in Fig. \ref{Figure3}(d).

\subsection{B$\textrm{a}_2$C$\textrm{o}$T$\textrm{e}_{1-x}$W$_x$O$_6$}

\begin{figure}
\centering
\includegraphics[width=1\linewidth]{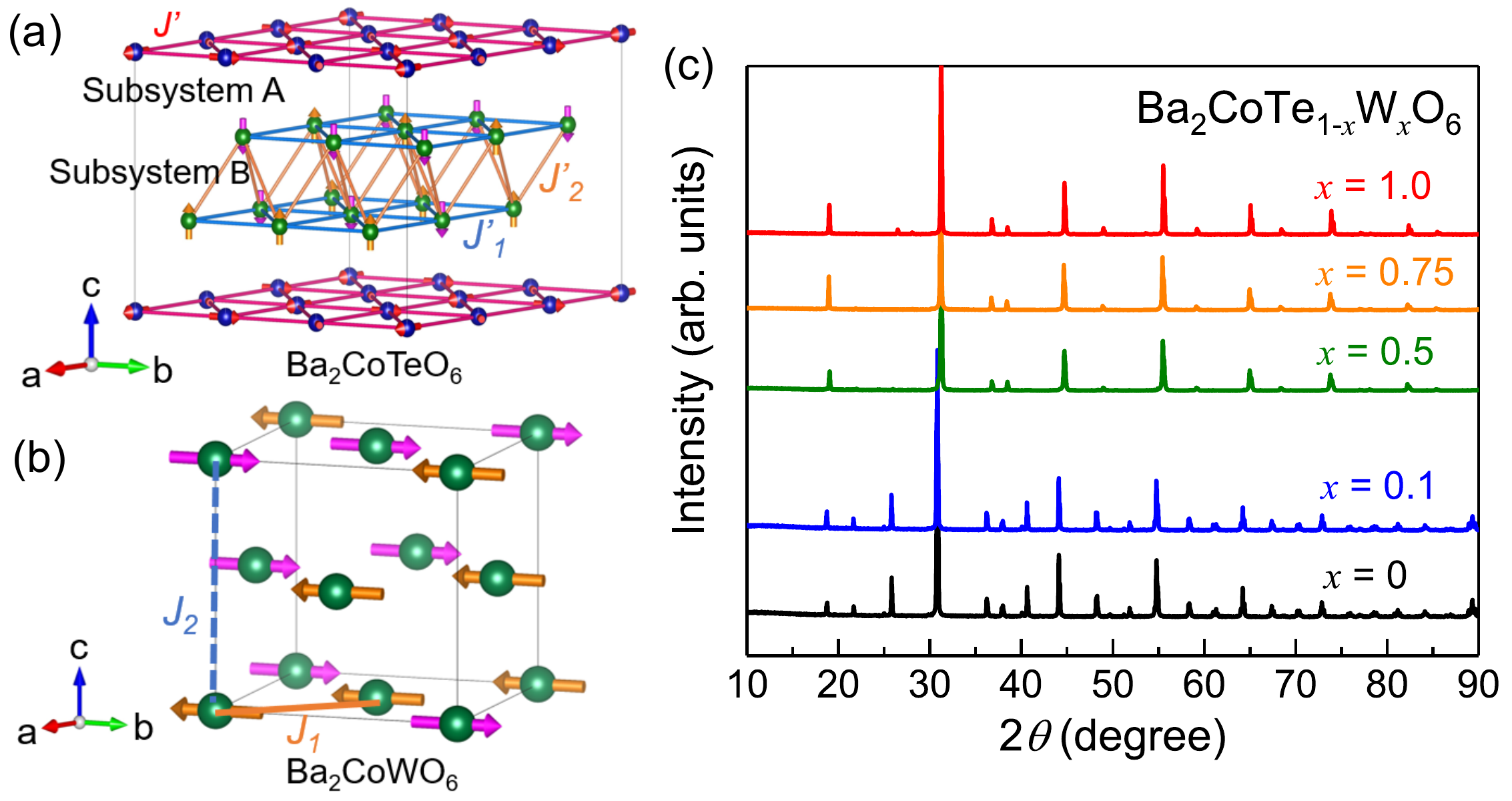}
\caption{(a) Crystal and magnetic structure of Ba$_2$CoTeO$_6$ only showing the Co$^{2+}$ ions arranging in a triangular lattice. There are two magnetic subsystems in Ba$_2$CoTeO$_6$, including the Heisenberg triangular lattice subsystem A with magnetic interactions $J'$, and the Ising buckled honeycomb subsystem B having intra-layer magnetic interaction $J_1'$ and inter-layer magnetic interaction $J_2'$. The magnetic ground state of subsystem A and subsystem B are a 120$^{\circ}$ spin structure and a stripy spin structure, respectively\cite{Ivanov2010,Chanlert2016,Chanlert2017,Kojima2022}. (b) Crystal and magnetic structure of Ba$_2$CoWO$_6$ only show Co$^{2+}$ ions arranging in a FCC lattice with NN interaction $J_1$ and NNN interaction $J_2$, whose magnetic ground state is AFM ordering type-II\cite{Cox1967}. (c) Doping dependence of XRD pattern for Ba$_2$CoTe$_{1-x}$W$_{x}$O$_6$.}
\label{Figure4}
\end{figure}

Figures \ref{Figure4} (a) and (b) depict the crystal structure of Ba$_2$CoTeO$_6$ and Ba$_2$CoWO$_6$, respectively. As mentioned above, Ba$_2$CoTeO$_6$ hosts a trigonal structure with space group $P\bar{3}m$, and the Co$^{2+}$ ions are arranged in a triangular lattice. Ba$_2$CoWO$_6$ crystallizes into a cubic structure with space group $Fm\bar{3}m$, and the Co$^{2+}$ ions construct an FCC lattice. The Co$^{2+}$ ions in these two compounds are surrounded by 6 O$^{2-}$ ions to form CoO$_{6}$ octahedra. The octahedral field and spin-orbit coupling together give rise to an $S_{eff}=1/2$ in low temperature for such a high-spin $S=3/2$ Co$^{2+}$ ion\cite{Chanlert2016,Chanlert2017,Liu2018,Yao2020}. Fig. \ref{Figure4}(c) displays the XRD patterns of Ba$_2$CoTe$_{1-x}$W$_{x}$O$_6$. The diffraction patterns for $0 \leq x \leq 0.1$ are quite different from that with $0.5 \leq x \leq 1$, which means different lattice structures. Specifically, compounds with $0 \leq x \leq 0.1$ maintain the trigonal structure (space group: $R\bar{3}m$), while compounds with $0.5 \leq x \leq 1$ hold the FCC structure (space group: $Fm\bar{3}m$). The refinements of compounds with $x$ = 0 and $x$ = 1.0 are shown in Appendix in Figs.~\ref{PXRD}(c) and (d), where tiny BaWO$_4$ impurity can be observed in Ba$_2$CoWO$_6$. It should be noted that there is a miscibility gap for the nominal doping range $0.1 < x < 0.5$, which makes the reaction products to be the mixtures of Ba$_2$CoTeO$_6$ and Ba$_2$CoWO$_6$.

\begin{figure} [t]
\centering
\includegraphics[width=1\linewidth]{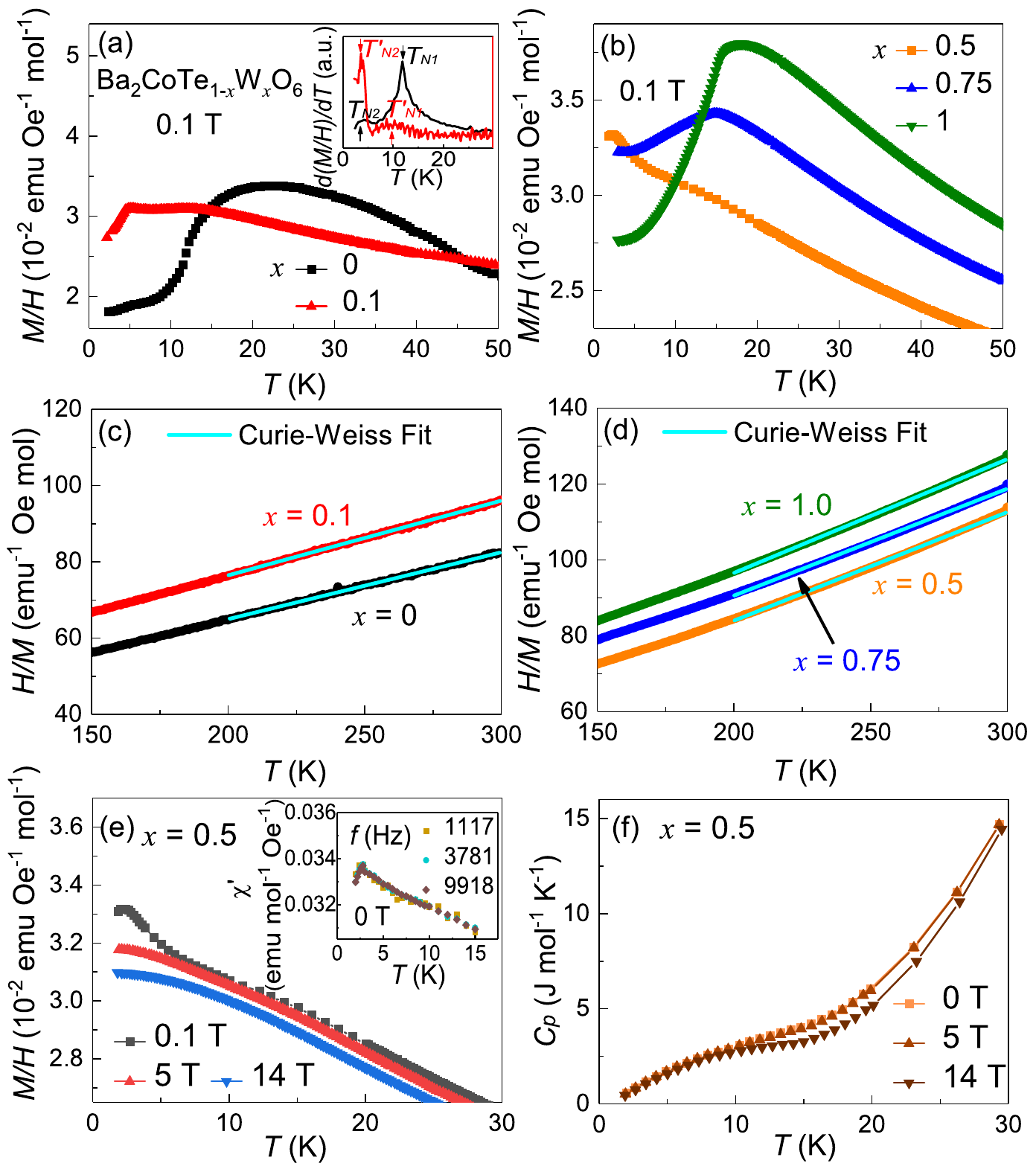}
\caption{(a) Magnetic susceptibility as a function of temperature of Ba$_2$CoTe$_{1-x}$W$_{x}$O$_6$ at doping $x$ = 0, 0.1, and (b) $x$ = 0.5, 0.75, and 1. The inset of (a) shows the derivative of susceptibility. (c)-(d) CW fittings of the DC susceptibility data of Ba$_2$CoTe$_{1-x}$W$_{x}$O$_6$ at temperatures above 200 K. These curves have been vertically shifted for clarity. (e) Temperature dependence of magnetic susceptibility of Ba$_2$CoTe$_{0.5}$W$_{0.5}$O$_6$. The inset shows the AC susceptibility at different frequencies. (f) Temperature dependence of heat capacity of Ba$_2$CoTe$_{0.5}$W$_{0.5}$O$_6$. }
\label{Figure5}
\end{figure}

Figure \ref{Figure5} shows the temperature dependence of DC magnetic susceptibility and heat capacity of Ba$_2$CoTe$_{1-x}$W$_x$O$_6$. Ba$_2$CoTeO$_6$ has two magnetic subsystems due to the distance difference between the adjacent triangular layers. These two decoupled magnetic subsystems give rise to two separate AFM transitions at $T_{N1}$ = 12 K and $T_{N2}$ = 3 K [Fig.~\ref{Figure5}(a)], corresponding to the AFM transitions of the magnetic subsystem B of buckled honeycomb lattice and the magnetic subsystem A with triangular lattice, respectively~\cite{Ivanov2010,Chanlert2016,Chanlert2017,Kojima2022}. Below $T_{N1}$, the subsystem B of Ising-like buckled honeycomb lattice enters a stripe-type AFM order (AFM$_1$) with magnetic the propagation vector $\bf{k}_1$ = (1/2, 0, 0), while the subsystem A of Heisenberg-like triangular lattice forms 120$^{\circ}$ AFM order (AFM$_2$) with the propagation vector $\bf{k}_2$ = (1/3, 1/3, 0) below $T_{N2}$, as shown in Fig.~\ref{Figure5}(a)~\cite{Chanlert2016,Chanlert2017,Kojima2022}. Compound with $x = 0.1$ also has two similar magnetic transitions at $T'_{N1}$ = 10 K and $T'_{N2}$ = 5.4 K [Fig.~\ref{Figure5}(a)], which should have the same mechanism with $x = 0$.

Ba$_2$CoWO$_6$ shows an AFM phase transition at $T_N$ $\sim$ 17 K in the DC susceptibility [Fig.~\ref{Figure5}(b)], which corresponds to a type-II order with the propagation vector $\bf{k}$ = (1/2, 1/2, 1/2)\cite{Cox1967}. Compound with $x=0.75$ has a similar peak at 15 K in susceptibility, which should also correspond to the transition to a type-II order similar to that in the Ba$_{2}$CoWO$_{6}$. As for the compound with $x=0.5$, there is a weak peak around 3 K in susceptibility, which will be discussed below. CW fittings above 200 K of the magnetic susceptibility of Ba$_2$CoTe$_{1-x}$W$_{x}$O$_6$ are shown in Figs.~\ref{Figure5}(c) and (d). The CW-fitted parameters, the deduced effective magnetic moment $\mu_{eff}$, and the frustration coefficient $f$ are summarized in Table \ref{Table2}. All the compounds have large negative CW temperatures, indicating dominant AFM interactions. As doping $x$ increases, the absolute value of $\mit\Theta_{CW}$ decreases, corresponding to weakening magnetic interactions with W$^{6+}$ doping. The effective magnetic moments are dramatically larger than $\mu_{eff}$ = $g\sqrt{S(S+1)}$ = 3.87 $\mu_B$ for a system with only $S$ = 3/2, which confirm the considerable spin-orbital coupling for Co$^{2+}$ ions in Ba$_2$CoTe$_{1-x}$W$_x$O$_6$. It should be noted that the magnetic parameters of spin-orbital-coupled systems determined from the CW law can be unreasonable~\cite{Li2021Modified}. For example, the obtained $\mit\Theta_{CW}$ and the frustration coefficient $f$ may be overestimated. The magnetic properties of the $x=0.5$ compound have been further studied in terms of magnetic susceptibility and heat capacity in the magnetic field [Figs. \ref{Figure5}(e) and (f)]. Higher magnetic fields can suppress the weak peak around 3 K in magnetic susceptibility. In heat capacity [Fig.~\ref{Figure5}(f)], there is no sharp peak around 3 K, but there is a broad feature below $\sim$10 K, which suggests that the weak peak around 3 K should correspond to a transition to a short-range order. As shown in the inset of Fig. \ref{Figure5}(e), the AC magnetic susceptibility of Ba$_2$CoTe$_{0.5}$W$_{0.5}$O$_6$, the weak peak near 3 K does not shift with frequency, which can rule out the spin-glass or spin-freezing magnetic ground state.
\begin{table}
\caption{Effective moments $\mu_{eff}$, Curie-Weiss temperatures $\mit\Theta_{CW}$, $T_{N}$, and frustration coefficient $f$ of Ba$_2$CoTe$_{1-x}$W${_x}$O$_6$ determined by CW fit.}
 \vspace{2mm}
\begin{tabular}{cccccccc}
\hline
$x$      & $\mu_{eff}$($\mu_B$) & $\mit\Theta_{CW}$(K) &$T_{N1}$(K) &$T_{N2}$(K) &$T_{N}$(K) &$f$ \\ \hline
0 & 6.71 & -170 & 12 & 3 &  & \\
0.1 & 6.35  & -139  & 10 & 3.6 &  & \\
0.5  &5.26  & -94 &  &  & 2.8 &33.6 \\
0.75 & 5.30  & -75 &  &  & 15.4 &4.9 \\
1.0  & 5.13  & -55 &  &  & 17.8  & 3.1\\   \hline \hline
\end{tabular}
\label{Table2}
\end{table}

\begin{figure}
\centering
\includegraphics[width=0.75\linewidth]{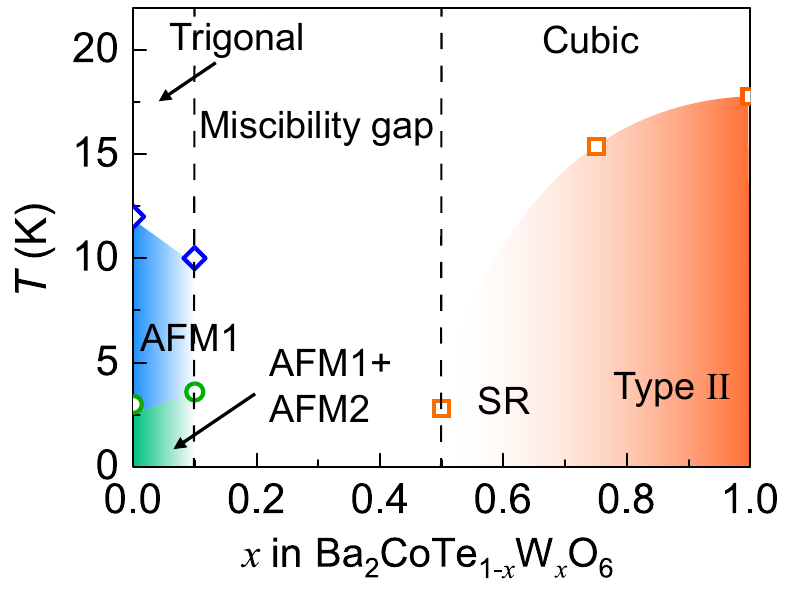}
\caption{ Doping and temperature magnetic phase diagram of Ba$_2$CoTe$_{1-x}$W${_x}$O$_6$. The blue diamonds, green circles, and orange squares indicate the transition temperatures of the AFM$_1$, AFM$_2$, and type-II order, respectively, determined by the magnetic susceptibility measurements. The miscibility gap is indicated between the two vertical dashed lines.}
\label{Figure6}
\end{figure}

Based on these results, a phase diagram concludes the crystal structure and magnetic ground state of Ba$_2$CoTe$_{1-x}$W$_x$O$_6$ is constructed, as shown in Fig. \ref{Figure6}. In the trigonal phase for $0 \leq x \leq 0.1$, the compounds have two phase transitions corresponding to two magnetic subsystems, including AFM$_1$ in $T_{N2}$ $\leq T \leq$ $T_{N1}$ and AFM$_2$ in $T \leq$ $T_{N2}$. The doping region has a miscibility gap of $0.1< x < 0.5$. In the cubic phase with $0.5 \leq x \leq 1$, AC susceptibility can rule out spin freezing, suggesting that the magnetic ground state evolves from a short-range order to the type-II long-range order.

\section{Discussion and conclusions}
The similar radius of W$^{6+}$ and Te$^{6+}$ ions make it easy to achieve chemical substitution. With the replacement of W$^{6+}$ ions to Te$^{6+}$ ions, the magnetic ground state of the double perovskite compounds Ba$_2$MnTe$_{1-x}$W$_x$O$_6$ are effectively changed without significant change on the crystal structure. We shall discuss the mechanism of such doping effect. As discussed in Refs. \cite{Babkevich2016,Mustonen2020}, the nonmagnetic Te$^{6+}$ and W$^{6+}$ cations act as the bridge between the magnetic Mn$^{2+}$ cations. The main differences of the magnetic interactions and ground states in the Te-based and W-based double perovskites can be explained as the differences of the orbital hybridization. In detail, the empty $5d$ orbital of W$^{6+}$ ion can hybridize with the O 2p orbital, thus providing a pathway for electron transfer for realizing the $J_2$ interaction, while the fully occupied $4d$ orbitals of Te$^{6+}$ ion are far below the Fermi level thus cannot participate this process for realizing the $J_2$ interaction.
As a result, the superexchange interactions are dominated by antiferromagnetic $J_1$ via $M^{2+}$-O$^{2-}$-O$^{2-}$-$M^{2+}$ paths and antiferromagnetic $J_2$ through $M^{2+}$-O$^{2-}$-W$^{6+}$-O$^{2-}$-$M^{2+}$ paths in the Te-based and W-based double perovskites, respectively\cite{Babkevich2016,Mustonen2020,Fogh2022}. As mentioned above, the magnetic ground states of the FCC Heisenberg model are determined by the $J_2/J_1$ ratio. The reported $J_2/J_1$ ratio in Ba$_2$MnTeO$_6$ (Ba$_2$MnWO$_6$) is $\sim$$-$0.09 ($\sim$0.95), which theoretically supports the type-I (type-II) AFM order and is consistent with the experimental results~\cite{Mustonen2020,Mutch2020}. Thus, The W$^{6+}$ and Te$^{6+}$ ions replacement in Ba$_2$MnTe$_{1-x}$W$_x$O$_6$ can change the relative strength of $J_1$ and $J_2$ and tune the magnetic ground state accordingly. However, the $J_2/J_1$ ratio change here may not be continuous since the predicted type-III AFM between type-I and type-II is not observed. As for the dip of CW temperature $\mit\Theta_{CW}$ in Fig.~\ref{Figure3}(c), it is possibly caused by an intermediate phase (for example, the random singlet state) that has no obvious phase transition signal in heat capacity results. A microscopic probe such as neutron scattering is highly required for uncovering the magnetic ground state at around $x=0.1$.

In Ba$_{2}$CoTe$_{1-x}$W$_{x}$O$_{6}$, Co$^{2+}$ ions host strong spin-orbital coupling and $S_{eff}=1/2$, which emphasizes the magnetic anisotropy and quantum effect. The FCC lattice retains in the doping region of $0.5 \leq x \leq 1$ with the type-II AFM order, which should follow the $J_1$-$J_2$ physics in the FCC Heisenberg model, as discussed above. Ba$_{2}$CoTeO$_{6}$ is an ideal platform to study the quantum mechanism that originates from both geometrical and interaction frustrations due to its two magnetic subsystems and has been well-studied before~\cite{Ivanov2010,Chanlert2016,Chanlert2017,Kojima2022}. Here, we provide the upper limit ($x$ = 0.1) of the W$^{6+}$ substitution to study such two-subsystem magnetism. Some candidate Kitaev magnets were proposed and well studied on cobalt-based honeycomb compounds including Na$_{2}$Co$_{2}$TeO$_{6}$, Na$_{3}$Co$_{2}$SbO$_{6}$ and BaCo$_{2}$(AsO$_{4}$)$_{2}$, etc.~\cite{VICIU2007,Bera2017,Yao2020,Zhong2020,Lin2021,Yao2022,Li2022,Fu2023,Zhang2023} It was proposed that the Kitaev model is also applicable in FCC lattice\cite{Kimchi2014}. Therefore, Ba$_{2}$CoTe$_{1-x}$W$_{x}$O$_{6}$ system could serve as a significant platform for studying the $J_1$-$J_2$ model and Kitaev physics.

In summary, this work presents the doping dependence of the crystal structure and magnetic ground state in magnetically frustrated systems Ba$_2$$M$Te$_{1-x}$W$_x$O$_6$ ($M$ = Mn, Co). For Ba$_2$MnTe$_{1-x}$W$_x$O$_6$, a classical magnetic system with $S$ = 5/2, the FCC crystal structure is not affected by the W$^{6+}$ doping, and the ground states are type-I AFM order when $0 \leq x \leq 0.08$ and type-II AFM order for $0.1\leq x \leq 1$, although the order states are short-range for some dopings. Ba$_2$CoTe$_{1-x}$W$_x$O$_6$ holds a trigonal lattice with two magnetic subsystems in the doping range $0\leq x \leq 0.1$, followed by a miscibility gap at $0.1< x < 0.5$. The compounds with $0.5\leq x < 1$ are in the FCC phase, where the ground state is type-II AFM order, which is the same as that in Ba$_{2}$MnWO$_{6}$. According to these results, the magnetic ground states of these frustrated systems have been successfully tuned by the Te$^{6+}$ and W$^{6+}$ substitution and the doping dependence of magnetic phase diagrams has been established. The different superexchange paths due to the differences of the orbital hybridization between the W$^{6+}$ and Te$^{6+}$ cases should be the mechanism for doping-mediated magnetic ground states in Ba$_2$$M$Te$_{1-x}$W$_x$O$_6$ ($M$ = Mn, Co).

\section*{Acknowledgements}

This work was supported by the National Key Research and Development Program of China (Grant Nos. 2023YFA1406500, 2024YFA1613100), the National Natural Science Foundation of China (Grant Nos. 12304187 and 12425404), the open research fund of Songshan Lake Materials Laboratory (Grant No. 2023SLABFN30), the Guangzhou Basic and Applied Basic Research Funds (Grant Nos. 2024A04J4024 and 2024A04J6417), Guangdong Provincial Key Laboratory of Magnetoelectric Physics and Devices (Grant No. 2022B1212010008), the Guangdong Basic and Applied Basic Research Funds (Grant No. 2024B1515020040), and Research Center for Magnetoelectric Physics of Guangdong Province (2024B0303390001).

\appendix
\section{Refinements on X-Ray Diffraction}
\begin{figure}[tbh]
\center{\includegraphics[width=\columnwidth]{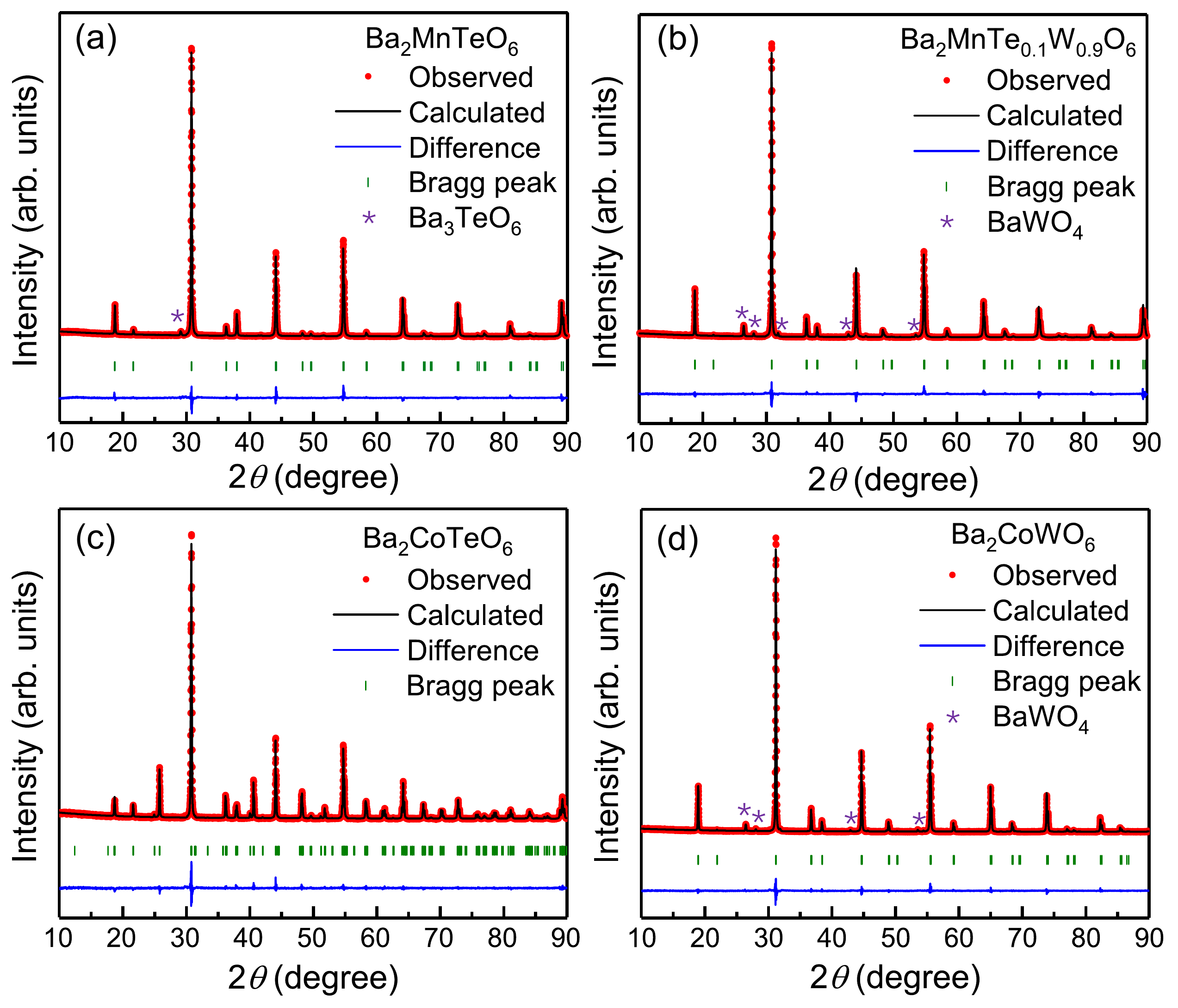}}
\caption{Rietveld refinements of four typical XRD data with different structures. The tiny peaks from the minor impurities are marked by asterisks (*).}
\label{PXRD}
\end{figure}

The structure of our sample are characterized by powder XRD using a Rigaku diffractometer. All the obtained results are refined with Rietveld method using $\textsc{FullProf}$ software~\cite{FullProf1993}. Some representative data and refinements are shown in Fig.\ref{PXRD}.

\end{document}